# Fabrication and electrical characterization of three-dimensional graphitic microchannels in single crystal diamond

**F Picollo[1,2], D Gatto Monticone[1,2, 3], P Olivero[1,2, 3], B A Fairchild[4], S Rubanov[5], S Prawer[4] and E Vittone[1,2, 3]**

E-mail: olivero@to.infn.it

[1]Experimental Physics Department and NIS Centre of Excellence, University of Torino, via P. Giuria 1, 10125 Torino, Italy

[2]Istituto Nazionale di Fisica Nucleare (INFN), Sezione di Torino, via P. Giuria 1, 10125 Torino, Italy

[3]Consorzio Nazionale Interuniversitario per le Scienze Fisiche della Materia (CNISM)

[4]School of Physics and Melbourne Materials Institute, University of Melbourne, Parkville, Victoria 3010, Australia

[5]Electron Microscope Unit, Bio21 Institute, The University of Melbourne, Parkville, Victoria 3010, Australia

**Abstract.** We report on the systematic characterization of conductive micro-channels fabricated in single-crystal diamond with direct ion microbeam writing. Focused high-energy (~MeV) helium ions are employed to selectively convert diamond with micrometric spatial accuracy to a stable graphitic phase upon thermal annealing, due to the induced structural damage occurring at the end-of-range. A variable-thickness mask allows the accurate modulation of the depth at which the microchannels are formed, from several µm deep up to the very surface of the sample. By means of cross-sectional transmission electron microscopy (TEM) we demonstrate that the technique allows the direct writing of amorphous (and graphitic, upon suitable thermal annealing) microstructures extending within the insulating diamond matrix in the three spatial directions, and in particular that buried channels embedded in a highly insulating matrix emerge and electrically connect to the sample surface at specific locations. Moreover, by means of electrical characterization both at room temperature and variable temperature, we investigate the conductivity and the charge-transport mechanisms of microchannels obtained by implantation at different ion fluences and after subsequent thermal processes, demonstrating that upon high-temperature annealing, the channels implanted above a critical damage density convert to a stable graphitic phase. These structures have significant impact for different applications, such as compact ionizing radiation detectors, dosimeters, bio-sensors and more generally diamond-based devices with buried three-dimensional all-carbon electrodes.

diamond/nanocarbon composites, 81.05.uj
        hopping transport, 72.20.Ee

Submitted to:    **New Journal of Physics**

## 1. Introduction

Diamond has attracted an ever-growing interest for the development of electronic devices with unprecedented performances [1], due to its extreme electrical properties such as high electron and hole mobility [2] and high electrical breakdown field ($10^7$ V cm$^{-1}$) coupled with extreme structural strength (Young modulus: E=1220 GPa), thermal conductivity (22 W cm$^{-1}$ K$^{-1}$) and broad electrochemical window. On the basis of such a promising potential, a significant effort has been made to optimize its interfacing with conventional electronics, and has motivated the development of techniques to fabricate electrical contacts and electrodes in diamond. Different approaches have been adopted, ranging from surface processing such as metallization [3] or hydrogen termination [4], to bulk doping achieved either by ion implantation [5] or CVD deposition [6]. Besides the above-mentioned techniques, the ion-beam-induced graphitization of diamond has been extensively investigated due to the availability of two stable structural forms for carbon at standard conditions of pressure and temperature, i.e. diamond and graphite [7]. The advantages of the latter approach consist in the possibility of obtaining conductive electrodes in an all-carbon environment, in the high thermal stability of the final structures, and in the possibility of defining contacts and electrodes into high-purity substrates without involving growth stages or complex lithographic processes. On the other hand, the main limitations consist in the need of ion beam facilities and high implantation fluences necessary to achieve the amorphisation of the material.

The process of graphitization induced by ion implantation was first reported in the 70s [8]. Following this pioneering study, the charge conduction mechanisms in damaged diamond have been investigated in detail for a broad range of ion species, energies (from few tens of keV up to MeV) and fluences. After the demonstration that the electrical properties of ion-implanted diamond layers are similar to those of amorphous carbon produced by sputtering graphite [9], the hopping conduction in diamond implanted with C ions was investigated [10]. In a series of Ar and C implantation experiments carried out at different temperatures, it was demonstrated that target temperature during implantation has a strong influence on the ion damage processes that determine the increase in conductivity [11]. The effect of ion implantation on the electrical properties was also studied in polycrystalline samples grown by chemical vapor deposition, demonstrating that the fluence dependence of the electrical conductivity of the implanted area is similar to what measured in single crystal samples [12, 13]. *In-situ* IV measurements were performed during ion implantation at variable temperatures, showing complex non-monotonic dependencies of the electrical conductivity as a function of the ion fluence [14]. Subsequently, an extensive IV characterization in temperature of diamond implanted with Xe and C ions at variable fluences while maintaining the sample at low temperature, allowed the extraction of important conduction parameters such as the characteristic energies for hopping sites [15, 16] and the densities of hopping centers [17].

The most widespread interpretation on the effects of ion-induced damage on the charge transport properties of diamond can be summarized as follows. As structural damage is introduced in the diamond lattice at increasing vacancy densities, a network of $sp^3$- and $sp^2$-bonded defects is progressively formed, which, above a certain critical level, is responsible for the amorphization of the material and therefore for the increase of its conductivity. Two models have been proposed to interpret the charge transport in this regime: variable range hopping [9, 10] and transport in a defect-related band [16]. The former model is based on the theory developed by Mott, in which conduction takes place via a temperature-dependent tunneling process between localized states around the Fermi level which are determined by randomly distributed defects in the lattice [18]. The latter model is based on the formation of a band of delocalized states within the forbidden gap, formed by the states associated to charged vacancy defects [19].

While the interpretation of the onset of hopping-related conduction at low damage densities has been debated in details in the above mentioned works, the conversion of the amorphized regions damaged above a critical threshold (often referred as "graphitization threshold") to a stable graphitic phase upon thermal annealing [20-26] is at the basis of the metallic conductivity of highly conductive layers in diamond.

The definition of graphitic conductive regions in diamond has found several interesting applications, such as: ohmic contacts [27-30], infrared radiation emitters [31], field emitters [32], bolometers in both single-crystal [33] and polycrystalline [34] samples, metallo-dielectric photonic crystals [35] and ionizing radiation detectors for X-ray [36] and MeV ion [37] beams.

In particular, the use of MeV ions allows the possibility of creating conductive graphitic micro-channels developing within the diamond bulk in the three spatial dimensions, with promising applications ranging from three-dimensional particle detectors to bio-sensors. While the definition of the structures in their lateral features is possible thanks to the employment of focused beams with ever-increasing spatial resolution [38], the peculiar damage profile of MeV ions in matter (see figure 1) allows high spatial resolution also in the depth direction.

The possibility of creating buried conductive structures in the diamond bulk with MeV ion implantation has not been extensively studied, particularly because it poses the challenge of developing a suitable strategy to implement their electrical connection with the sample surface at specific locations. Different methods have been explored to address this issue, such as laser-induced graphitization [39], high-voltage-induced thermal breakdown [40], multiple-energy ion implantation [34] and laser ablation [41]. In a preliminary study [42], we proposed the use of three-dimensional masks to modulate the penetration of MeV ions in diamond, thus allowing the direct and fast writing of three-dimensional channels in the material with high spatial accuracy. Following the subsequent studies on the conduction mechanisms occurring in MeV-implanted diamond [43, 44], in the present paper we present a detailed analysis of the structural features of three-dimensional graphitic micro-channels by means of cross sectional TEM. Moreover, measurements of the temperature-dependent electrical properties of the channels implanted at various fluences and after different annealing stages shows conclusively that micro-channels with an electrical conductivity equivalent to that of polycrystalline graphite can be defined in the three spatial dimensions in a fully insulating single-crystal diamond matrix.

**2. Experimental**

*2.1. Sample masking for ion implantation*

In the present work three samples were employed. Sample 1 consists of a single-crystal diamond produced by Sumitomo Electrics with High Pressure High Temperature (HPHT) technique; it is classified as type Ib with a substitutional nitrogen concentration of ~10-100 ppm, and it was employed for structural characterization with cross-sectional Transmission Electron Microscopy (TEM). Samples 2 and 3 consist of single-crystal diamonds produced by ElementSix with Chemical Vapour Deposition (CVD) technique; they are classified as type IIa with substitutional nitrogen and boron concentrations of <1 ppm and <0.05 ppm, respectively. They were employed for electrical characterization at various temperatures. All samples are ~3×3×0.5 $mm^3$ in size and their main faces are oriented along the (100) crystallographic direction.

All samples were implanted with a raster-scanning MeV He microbeam at variable fluences to define buried micro-channels at a depth of few μm from the surface. All implantations were performed at room temperature using currents < 30 nA to minimize sample heating and so as to avoid *in situ* graphitization [14]. As shown in figure 1, the damage density induced in the diamond structure by He ions with energies of 0.5 MeV, 1.3 MeV and 1.8 MeV has a peculiar depth profile with a pronounced peak at their end of range depths, corresponding respectively to ~0.96 μm, ~2.20 μm and ~3.3 μm. The profiles were obtained with the SRIM-2008.04 Monte Carlo code [45] in "Detailed calculation with full damage cascades" mode, by setting a displacement energy value of 50 eV [46]. As mentioned above, a variable-thickness masking system has been optimized in order to modulate the penetration range of implanted ions, and therefore control with high spatial accuracy the depth at which the material is heavily damaged. Variable-thickness contact masks were defined on all samples prior to ion implantation by means of a thermal evaporation of metal through apertures with dimensions of ~250×90 µm µm kept at a finite distance (typically ~300 µm) from the samples surfaces, thus taking advantage of a typical "shadow evaporation" mechanism [47]. Ag masks were evaporated on samples 1 and 3, while Cu was employed for sample 2. As shown schematically in figure 2a, the evaporation through a non-contact aperture allows the definition on the sample surface of contact metal masks with pre-defined shapes and sizes, characterized by a constant thickness in their central regions and with smooth and slowly degrading edges at their respective corners. The latter features are determined by a "shadow" effect during the evaporation process which can be suitably controlled by setting the sample-to-aperture distance. As shown in figure 2b, the MeV ion implantation across such edges determines the formation of highly damaged layer at a modulated depth, thus allowing the connection of the endpoint of the buried layer with the sample surface. The method described in this work represents a significant improvement with respect to that reported in earlier works [42-44], allowing a reliable and reproducible lithographic process by means of variable-thickness masks with smoothly sloping edges, as reported in the profilometry scan reported in figure 2c.

*2.2. TEM characterization*

In order to perform a cross-sectional TEM characterization of the buried channels, sample 1 was implanted at room temperature with 0.5 MeV $He^+$ ions at the MP2 microbeam line of the 5U NEC Pelletron accelerator of the University of Melbourne. A region of $100 \times 1220$ $\mu m^2$ was implanted at a fluence of $1 \cdot 10^{17}$ $cm^{-2}$ across the sloping edge of a 1.2 µm thick Ag mask. The focused ion microbeam was scanned across the irradiated area to deliver a homogeneous ion fluence; the ion beam current was ~10 nA, thus determining an implantation time of ~30 min.

After ion implantation, a thin cross section of the sample was cut with Focused Ion Beam (FIB) milling along the <110> crystallographic direction, i.e. at an angle of 45° with respect to the channel's longitudinal axis, in order to directly image the emerging profile of the buried amorphous layer formed at the ion end of range. The FIB used was a dual-beam FEI Nova Nanolab 200 instrument equipped with micro-manipulator for *in situ* TEM sample preparation. After ion milling, the thin cross section was disconnected from the sample and fixed on a TEM grid by means of a micro-manipulator, following standard procedures for FIB-based preparation of TEM samples [25]. Cross sectional TEM images were acquired using Tecnai TF20 electron microscope operated at 200 keV.

In figure 3 a cross-sectional TEM image of the implanted sample along the sloping edge of the variable-thickness mask is shown. The amorphous damage layer is clearly visible in figure 3 due to the absence of any diffraction contrast. Note that this sample has not been annealed, and the end of range damage is in the form of amorphous carbon, rather than polycrystalline graphite. Consistent with what is predicted by SRIM simulations, the buried amorphized layer is visible at a depth of ~1 μm below the sample surface and its progressive and continuous emergence towards the sample surface follows the profile of the variable-thickness mask from the left-hand to the right-hand side of the figure. TEM cross-sectional imaging provides a direct evidence of the continuous connection of buried amorphized layers with the sample surface, as confirmed by the results of the electrical characterization described in the following section.

*2.3. Electrical characterization*

With the purpose of creating conductive channels for electrical testing, samples 2 and 3 were implanted with $He^+$ ions with energies 1.3 MeV and 1.8 MeV, respectively. The implantations were performed at the ion microbeam line of the AN2000 accelerator at the Legnaro National Laboratories of the Italian National Institute of Nuclear Physics (INFN). Also in this case a raster-scanning ion microbeam was employed to optimize the resolution and uniformity of the implanted regions. Ion currents varied between 3 nA and 5 nA, thus allowing implantations to be performed in typical times of 10 mins. In both samples 7 micro-channels were implanted at fluences comprised between $2·10^{16}$ cm$^{-2}$ and $5·10^{17}$ ions.cm$^{-2}$. In sample 2 the channels had lengths between 985 μm and 1020 μm, while their widths were comprised between 15 μm and 25 μm. In sample 3, the channels lengths and widths were 415-425 μm and 19-25 μm, respectively. For each channel, the implantation was performed between the two sloping edges of the above described variable-thickness masks, so that the channels endpoints would connect to the sample surface, as schematically shown in figure 4a. After implantation, the variable-thickness masks were removed and the sample surfaces were cleaned in acetone ultrasonic bath, taking advantage of the poor adhesion of the evaporated metal on the polished diamond surface. In figure 4b, the optical image in transmission of two typical channels implanted at fluences $7.7·10^{16}$ cm$^{-2}$ and $1.6·10^{17}$ cm$^{-2}$ in sample 2 is shown after removal of the mask: the channels are clearly visible since the amorphous phase is optically opaque with respect to the surrounding diamond matrix. The channels were subsequently contacted by evaporating Ag pads at their endpoints and also in intermediate positions along the channels extension, as shown schematically in figure 4c, in which the contact pads are labeled as A and D. The electrical characterization of the channels at variable temperature was performed with a home-built setup consisting in a system of microprobes controlled by a set of micromanipulators operating in a vacuum chamber; the microprobes could contact different pads while the sample temperature was ramped from room temperature to 500 °C, with the sample temperature being measured to an accuracy of ~5 °C. The microprobes were connected with suitably shielded feedthroughs to a Keithley 2636 electrometer, allowing the acquisition of current-voltage (I-V) characteristics in a two-terminal configuration at variable temperature from different contacts on channels implanted in the above-mentioned fluences range. The availability of electrical contacts in intermediate positions (B, C in figure 4c) allowed the electrical probing of the sample at locations where the channels are not expected to be in contact with the surface. The process was repeated after annealing the samples in a range of temperatures (i.e. 200 °C and 400 °C for sample 2, 700 °C and 1000 °C for sample 3). For each of the above-mentioned steps (200 °C, 400 °C, 700 °C, 1000 °C), samples were annealed for 2 hours in an oxygen-free environment, in order to avoid surface etching.

*2.3.1. Characterization at room temperature.* The I-V characteristics obtained at room temperature from as-implanted sample 2 by connecting the sensing probes to contacts A-D (see figure 4c) are reported in figure 5, while similar trends (not reported here) are obtained for sample 3. The different I-V curves are relevant to the 7 channels implanted at increasing fluences (the exact values are reported in the legend of figure 5a). As shown in the linear-linear plot reported in figure 5a, the curves exhibit a quasi-linear trend in the (-10 V, +10 V) interval. Moreover, as shown more clearly in the linear-logarithmic plot of figure 5b, at any given voltage the current monotonically increases with the implantation fluence, as expected from channels with increasing damage densities and thicknesses.

The same measurements performed on the intermediate channels B and C (see figure 4c) are reported in figure 6. For each channel, the current values for a given voltage are 2-3 orders of magnitude lower with respect to the previous probing connections. This indicates that, although different paths are accessible in parallel for carriers along the implanted regions, the conduction mechanism is dominated by charge transport occurring in the buried heavily-damaged layers. Nonetheless, also in this case, a finite conductance is measured with the same monotonic correlation with increasing implantation fluence. This result indicates that the defect density induced by ion damage in the "cap layer" comprised between the heavily damaged buried layer and the sample surface, although significantly lower than in the buried channels, is high enough to induce a marginal increase of conductance in the material [48]. As expected, such an increase is more pronounced for regions implanted at higher fluences, while for the channel implanted at the lowest fluence (#7) the change in conductance is not measurable within the electrometer sensitivity. Also in this case, similar trends (not reported here) are obtained for sample 3. It is worth stressing that prior to ion implantation no conductance could be measured within the experimental sensitivity between test surface electrodes, i.e. the conductance of pre-implanted diamond is virtually zero for the purposes of the present investigation, and the contribution to conductance through surface states or residual contaminations are negligible with respect to the effects of ion implantation.

The effect of annealing on the channels conductance was investigated by monitoring the room-temperature I-V characteristics of the channels in both of the above-mentioned contact configurations at different stages of thermal processing. As an example, figure 7 reports the I-V curves for the channel implanted at fluence $F=4.5 \cdot 10^{17}$ cm$^{-2}$ in sample 2, after annealing at 200 °C and 400 °C, as described above. As shown in figure 7a, the conductance between contacts A and D (see figure 4c) monotonically increases with increasing annealing temperature, thus demonstrating that the effect of the thermal processing is to progressively "graphitize" (i.e. to convert to a stable $sp^2$ graphite-like phase) the buried amorphized channels created by ion implantation. On the other hand, as shown in figure 7b the conductance between contacts B and C (see figure 4c) exhibits a decreasing trend with the progressive annealing stages, indicating that within the low-damaged "cap layer" the effect of the thermal processing is to progressively recover the pristine diamond structure instead of inducing the graphitization process. It is worth noting that the latter trend is observed also for the conductance between contacts A and D for channels implanted at low fluence.

The variation in the channel resistance for channels implanted at different fluences was investigated systematically after each annealing step for both samples 2 and 3 and processed at different annealing temperatures.

In order to extract the physical parameters of the ion induced damage and independent on the irradiation patterns, we adopt the concept of sheet resistance $R_{sh}$, borrowed from the well known procedure to characterize integrated circuit resistors [49]. The sheet resistance is defined through the definition of the total conductance G of the entire implated region of width w and length L:

$$G = \frac{1}{R} = \frac{w}{L} \cdot \int_0^Z \sigma(z) dz = \frac{w}{L} \cdot \frac{1}{R_{sh}} \tag{1}$$

where $R$ is the resistance between contacts A and D, and σ is the local conductivity at a depth z from the surface; the integral is extended from the surface ($z=0$) down to the penetration depth ions ($z=Z$). From (1) follows that the sheet resistance:

$$R_{sh} = \frac{w}{L} \cdot R = \frac{1}{\int_0^Z \sigma(z) dz} \tag{2}$$

is defined as the resistance of a square implantation area ($w=L$) and, as such, conventionally specified in ohms per square.

If we assume a linear dependence of the local conductivity from the local density of vacancies ν(z) induced by ions (i.e. $\sigma(z) = C \cdot \nu(z) = C \cdot p(z) \cdot F$, where $C$ is a proportionality factor, $p(z)$ is the linear density of vacancies reported in the plots of figure 1 and $F$ is the ion fluence), the sheet resistance turns out to be inversely proportional to the damage surface density (expressed in number of vacancies per surface unit) defined as the product of the total number of vacancies $V$ generated by a single ion (evaluated by integrating along the depth direction the damage density profiles $p(z)$ reported in figure 1) and the fluence:

$$R_{sh} = \frac{1}{C \cdot V \cdot F} \tag{3}$$

It is worth noticing that (3) has been obtained by modeling the effect of cumulative ion implantation on the number of vacancies with a simple linear approximation which does not take into account more complex processes such as self-annealing, ballistic annealing and defect interaction and assuming a linear dependence of the conductivity on the vacancy density.

Nonetheless, the surface damage density can be regarded as a useful first-order parameter to re-scale fluences relevant to implantations performed with different ion species and/or energies.

In figures 8a and 8b the sheet resistance of channels implanted at increasing fluences is reported as a function of the damage surface density for samples 2 and 3, respectively. Different curves are reported for as-implanted channels and for the after-annealing cases. The IV characteristics measured between contacts A and D display ohmic behaviour for all implantation fluences and, as expected, the channels sheet resistance monotonically decreases with increasing damage densities. On the other hand, for channels implanted above a given damage surface density the effect of thermal annealing is to progressively decrease their sheet resistance: this evidence is compatible with a process of progressive graphitization of the amorphized buried layers. At increasing annealing temperatures, the sheet resistance of channels implanted at the highest fluences tends to saturate to a value of ~50 Ω/□ upon annealing at the highest temperature of 1000 °C, indicating that the graphitization process is nearly completed. As mentioned above, in channels implanted below a critical surface damage density the thermal annealing has the opposite effect of progressively "healing" the pristine diamond structure and increasing the measured resistance to values approaching the inner resistance of the electrometer, i.e. restoring the material to its pre-implantation condition, when a virtually null conductivity was measured, as mentioned above. As reported in figure 8, the threshold damage surface density can be roughly estimated as ~2·10$^{18}$ vacancies cm$^{-2}$. As previously mentioned, the damage surface density is defined as $\sigma_V = V \times F$, where $V$ is the number of vacancies created per incoming ion and $F$ is the fluence; on the other hand, the local vacancy density can be estimated in the same first-order approximation as $v(z) = p(z) \times F$, where $z$ is the depth in the material and $p(z)$ is the linear density of vacancies reported in the plots of figure 1. Therefore, the above-mentioned threshold value for the damage surface density corresponds to a peak vacancy density of $\nu(z^*) = p(z^*) \times \frac{\sigma_V}{V} \cong 8 \cdot 10^{22}$ vac cm$^{-3}$, where $z^*$ is the ion-end-of-range depth and the values $p(z^*) \cong 2 \cdot 10^{6}$ vac ion$^{-1}$ cm$^{-1}$ (see figure 1) and $V \cong 50$ vac ion$^{-1}$ are used. The obtained value represents an estimation of the critical vacancy density above which the damaged diamond structure converts to a stable graphite-like phase upon thermal annealing, a quantity which is often referred as "amorphization threshold" or $D_c$ in literature. Although obtained with the above-mentioned simplicistic linear approximations, our estimation of the amorphization threshold is in good agreement with the values of $9 \cdot 10^{22}$ vac cm$^{-3}$ [50], $7 \cdot 10^{22}$ vac cm$^{-3}$ [17], $6 \cdot 10^{22}$ vac cm$^{-3}$ [26] and $5 \cdot 10^{22}$ vac cm$^{-3}$ [51] available in literature.

The value of the amorphization threshold obtained above can be used to estimate the thickness of the channel, by evaluating the intersection of such a threshold with the vacancy density depth profile; the resulting estimation for the channel implanted at the highest fluence ($5.1 \cdot 10^{17}$ ions·cm$^{-2}$) is $t \cong 550$ nm. As reported in figure 8, the sheet resistance of the channel implanted at the highest fluence in sample 3 after high temperature annealing is $R_{sh} \cong 50$ Ω/□, therefore its resistivity can be estimated as $\rho = R_{sh} \times t \cong 2.8 \cdot 10^{-3}$ Ω cm, a value which is comprised between the resistivities along the directions which are respectively parallel ($\rho \cong 4.0 \cdot 10^{-5}$ Ω cm) and perpendicular ($\rho \cong 1.5 \cdot 10^{-1}$ Ω cm) to the layer planes in highly oriented pyrolytic graphite, and is in good agreement with the typical resistivity values measured for standard polycrystalline graphite ($\rho \cong 3.5 \cdot 10^{-3}$ Ω cm) [52].

On the other hand, it is worth remarking that after the final high-temperature annealing no residual conductivity of the channels between contacts B and C could be measured within the sensitivity limits of the employed setup, demonstrating that MeV ion implantation with a focused beam coupled with variable-thickness masking is a reliable method to create highly conductive microchannels embedded in a highly insulating diamond matrix.

*2.3.2. Characterization at variable temperature after 200 °C annealing.* Current-voltage characterization was also performed at variable temperature, with the purpose of elucidating the nature of the charge transport mechanisms within the microchannels for different damage densities before full graphitization.

The channels of sample 2 were investigated after 200 °C annealing, in order to avoid *in situ* annealing processes during the measurements at temperatures comprised between 25 °C and 160 °C: the adoption of a rather restrict temperature range was again motivated by the necessity of avoiding dynamic annealing processes during the measurement. Figure 9 reports as an example the resulting measurements of the resistance (*R*) at different temperatures (*T*) from the channel implanted at fluence $4.8 \cdot 10^{17}$ ions.cm$^{-2}$, both between contacts A-D (see figure 9a) and between contacts B-C (see figure 9b). The data are represented by plotting *ln(R)* versus $T^{-1/4}$, consistently with Mott's law describing the three-dimensional variable range hopping process:

$$R(T) = R_0 \cdot \exp\left(\frac{T_0}{T}\right)^{1/4} \tag{4}$$

where $R_0$ is the temperature-independent pre-factor.

The temperature $T_0$ introduced in (4) is defined as:

$$T_0 = \frac{512}{9\pi} \cdot \frac{\alpha^3}{n(E_F) \cdot k_B} \tag{5}$$

where α is the inverse of the radius of the localized wavefunction of the defect states that mediate the hopping conduction, $n(E_F)$ is the density of localized states at the Fermi level per unit of volume and energy, and $k_B$ is the Boltzmann constant.

As shown in figure 9, the experimental data are compatible with the "variable range hopping" model for both the A-D and B-C contact configurations, although it should be remarked that the rather restricted temperature range (motivated by the necessity of avoiding *in situ* annealing processes, as mentioned above) does not allow a fully unequivocal attribution of the charge transport mechanism and experimental data exhibit some compatibility also with the defect-related-band model (see below). Eqs. (1) and (2) can be employed to evaluate the value of the density of localized states at the Fermi level $n(E_F)$ from the slope of the fitting linear curves, by employing a value of $\alpha^{-1} = 1.2$ nm, as derived from a study of sputtered carbon films [53]. The plot in figure 10 reports the resulting $n(E_F)$ values for channels implanted at increasing fluences, for both the A-D and the B-C contact configurations. As expected, the density of localized states at the Fermi level monotonically increases of several orders of magnitude with increasing damage densities, while being significantly higher for the buried layer with respect to the cap layer.

*2.3.3. Characterization at variable temperature after 400 °C annealing.* The channels of sample 2 were electrically characterized in the A-D contact configuration in a broader range of temperatures after annealing at 400 °C. Measurements in the B-D contact configuration did not yield indicative results because the conductivity of the cap layers could not be measured within the instrument sensitivity after the annealing process (see also the following section). In A-D contact configuration, different trends of the resistance vs temperature were observed, depending significantly from the implantation fluence.

For the channels implanted at the highest fluences (i.e. above $3 \cdot 10^{17}$ ions.cm$^{-2}$), neither the "variable range hopping" model nor the "defect-related band" models are compatible with experimental data. This is because the observed "$R$ vs $T$" curves are approaching the trends characteristic of graphite, which will be presented more consistently in the following section (i.e. after full graphitization occurring with 1000 °C annealing).

At intermediate fluences (i.e. ~$3 \cdot 10^{17}$ ions.cm$^{-2}$), the observed trends are still compatible with the "variable range hopping" model, while exhibiting a less satisfactory consistency with the "defect-related band" model, which predicts an Arrhenius-type exponential dependence of resistance from temperature:

$$R(T) = R_0 \cdot \exp\left(+\frac{E_a}{k_B \cdot T}\right) \quad (6)$$

where $E_a$ is the activation energy for free carriers to be de-localized in the conduction sub-band [19].

Remarkably, at lower implantation fluences the observed trend exhibit a more satisfactory consistency with the "defect-related band" model with respect to the "variable range hopping" model. Although a complete investigation on the conduction mechanisms in amorphized diamond is beyond the scopes of the present work, these results seem to indicate that no conclusive model can be employed to satisfactorily model the charge transport mechanisms in sub-superficial amorphized layers for all implantation fluences. This fact can be attributed to the formation of a heterogeneous phase with graphitic crystals embedded in amorphous carbon, particularly at medium fluences and/or intermediate annealing processes.

*2.3.4. Characterization at variable temperature after 1000 °C annealing.* Sample 3 was characterized after 1000 °C annealing, in order to explore the temperature-dependent conduction mechanism within the graphitized microchannels. Figure 11 shows the variation of the resistance between contacts A and D for the channel implanted at a fluence of $2.5 \cdot 10^{17}$ ions·cm$^{-2}$ as a function of measurement temperature. The resistance is normalized to $R_0$, i.e. the extrapolated value at 0 °C, in order to compare the measured values with the empirical polynomial function describing the dependence of $R/R_0$ vs $T$ in common graphitic carbon, as reported in [54]:

$$\frac{R(T)}{R_0} = 1 - 3.17 \cdot 10^{-4} \cdot T + 1.73 \cdot 10^{-7} \cdot T^2 - 6.3 \cdot 10^{-11} \cdot T^3 \quad (7)$$

where the temperature $T$ is expressed in °C. As mentioned in the previous section, it is worth stressing that the trend reported in (7) is not compatible with either Mott- or Arrhenius-like trends.

The good agreement between experimental data and the phenomenological curve reported in [54] confirms that the conduction mechanism in the high-temperature-annealed microchannels is the same as in common polycrystalline graphite. It is worth noting that, for the channels implanted at the highest fluences ($F > 4 \cdot 10^{17}$ cm$^{-2}$) a graphitic-like conduction trend vs temperature was already achieved after 800 °C (data not reported here). As mentioned above, measurement in the B-C contact configuration could not be carried since there was no residual conductivity left in the cap layers after the high-temperature annealing process, as measurable within the sensitivity of the employed electrometer.

## 3. Conclusions

In the present work, a method for the direct ion beam writing of buried conductive micro-channels in diamond has been described, in which variable-thickness masks are employed to modulate the ion penetration depth and therefore define the shape of the microstructures in the depth direction, allowing in particular the emergence of the micro-channels at specific locations of the sample surface.

Cross-sectional TEM microscopy has been employed to directly image the emerging amorphized layers which are formed upon the damage-induced conversion of the diamond lattice to a stable $sp^2$ phase, demonstrating the reliability of the variable-thickness masking technique, which has been optimized with respect to previous reports [42-44] to define amorphous structures in diamond with micrometric resolution in the three spatial dimensions.

Direct electrical probing performed both at room temperature and at variable temperature on as-implanted and annealed samples indicated that:

- in the as-implanted samples, channels implanted at increasing fluence exhibit higher conductance when probed at their ending points, while similarly increasing but significantly lower conductance values are measured *above* the channels, i.e. in electrical contact with the "cap layer" extending between the buried micro-channels and the sample surface, indicating residual damage present in the as implanted cap layer;

- thermal annealing has the effect of progressively increasing the conductance of the micro-channels implanted above a critical threshold, while for channels implanted below such a threshold the conductance progressively decreases until, for high-temperature annealing, the cap layer recovers its pristine highly insulating behaviour;

- the above-mentioned threshold damage density has been roughly estimated to be ~$8·10^{22}$ vacancies $cm^{-3}$, in satisfactory agreement with values reported in previous reports [17, 26, 50];

- before high temperature annealing, the conduction mechanism within the micro-channels can be suitably described with the "variable range hopping" model, yielding increasing values of the density of the localized states near the Fermi level for increasing implantation fluences;

- after medium-low temperature annealing (200 °C, 400 °C), the channels implanted at the highest fluences start to exhibit evidence of the initial stages of graphitization, while channels implanted at intermediate fluences exhibit different degrees of compatibility with the "variable range hopping" and "defect-related band" models depending on the implantation fluences; more systematic studies are necessary to shed further light on this aspect;

- after high temperature annealing, besides the above-mentioned full recovery of the cap layer to the pristine values of resistivity, the buried amorphized layers fully convert to a graphitic phase, as demonstrated by the saturation of the conductance values for channels implanted at different fluences, and by the trend of the resistivity vs measurement temperature, in good agreement with what reported in literature for polycrystalline graphite [54].

In conclusion, the technique is demonstrated to be a versatile tool in the definition of buried micro-electrodes in diamond in a full-carbon environment, with promising applications in ionizing radiation detection [37] and bio-sensing, which will be explored in forthcoming works.

**Acknowledgments**

The work of P. Olivero is supported by the ''Accademia Nazionale dei Lincei –Compagnia di San Paolo'' Nanotechnology grant, by the "Dia.Fab." Experiment at the INFN Legnaro National Laboratories and by the FIRB 2010 project "RBFR10UAUV", which are gratefully acknowledged.

A. Alves and V. Rigato are gratefully acknowledged for their assistance at the ion microbeam facilities of the University of Melbourne and of the INFN Legnaro National Laboratories, respectively.

**Figures**

Figure 1

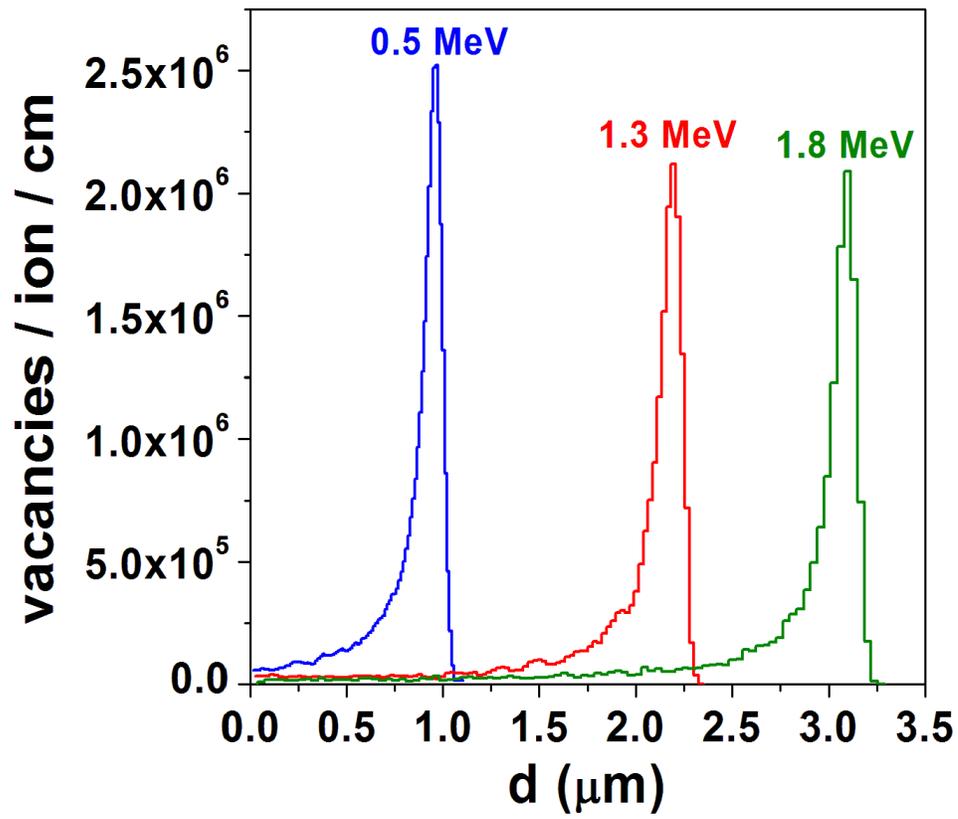

Figure 2

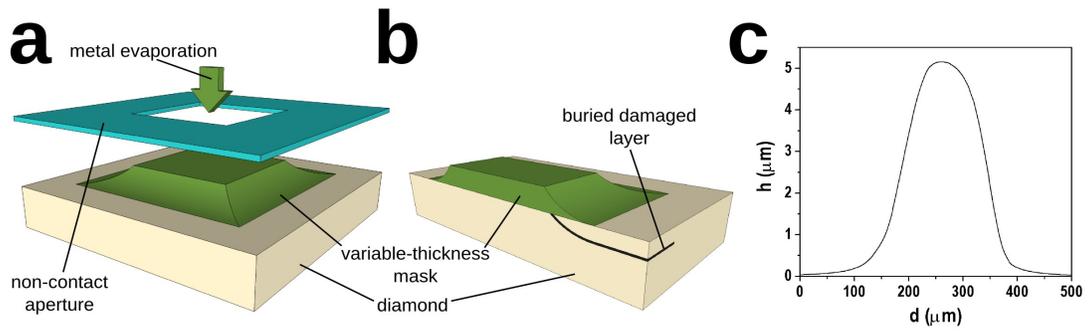

Figure 3

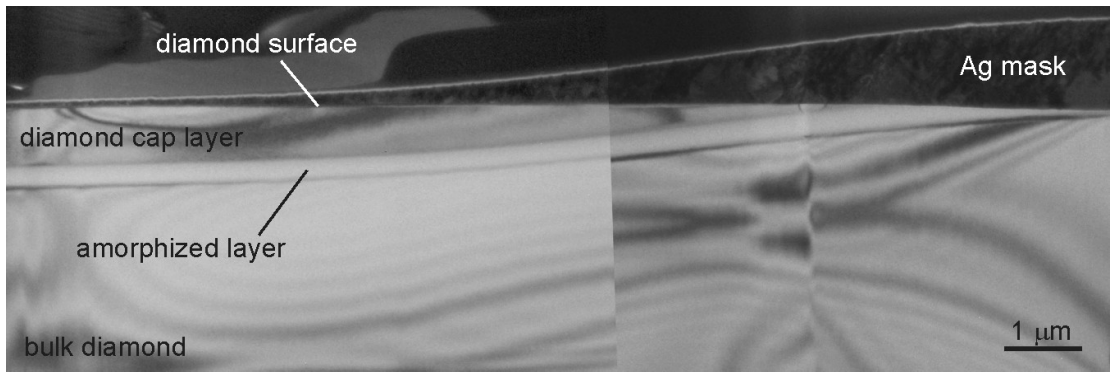

Figure 4

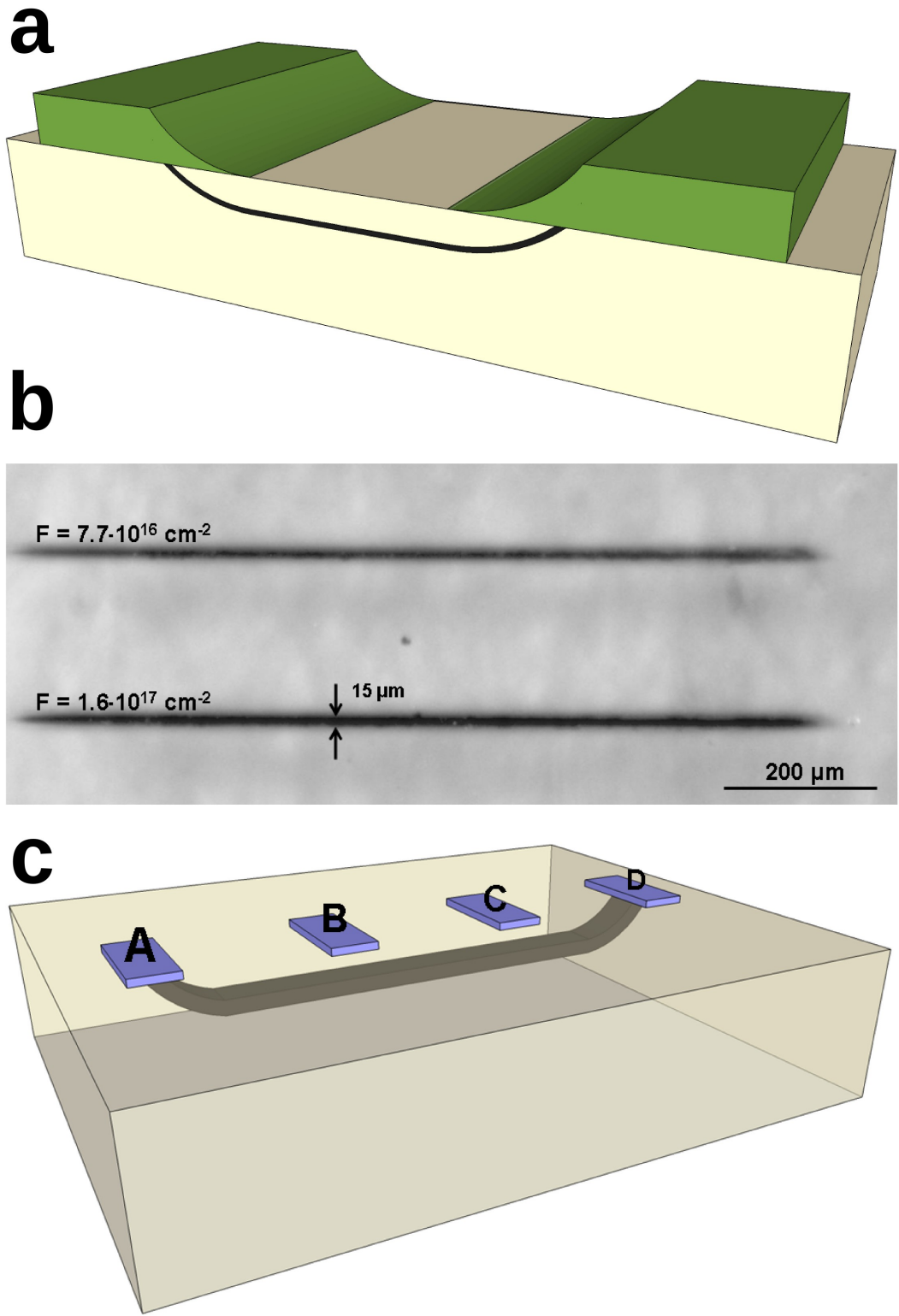

Figure 5

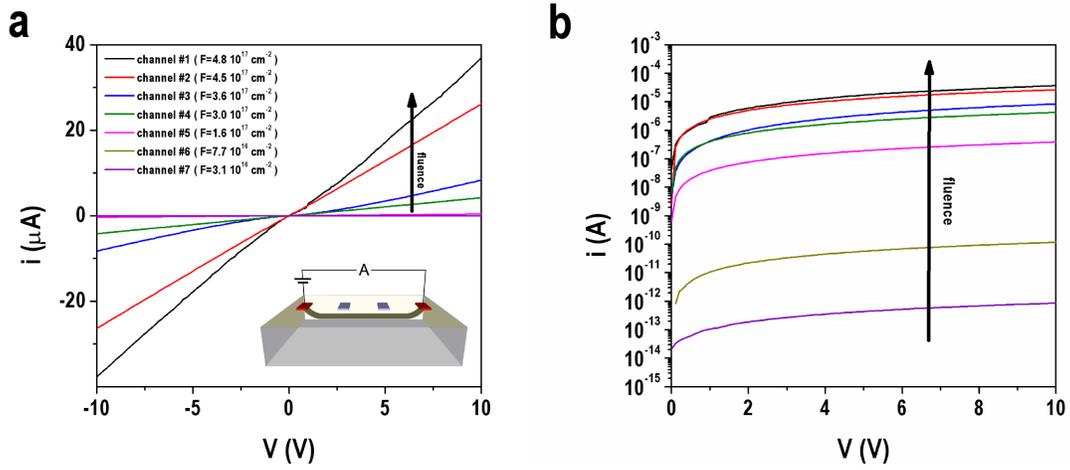

Figure 6

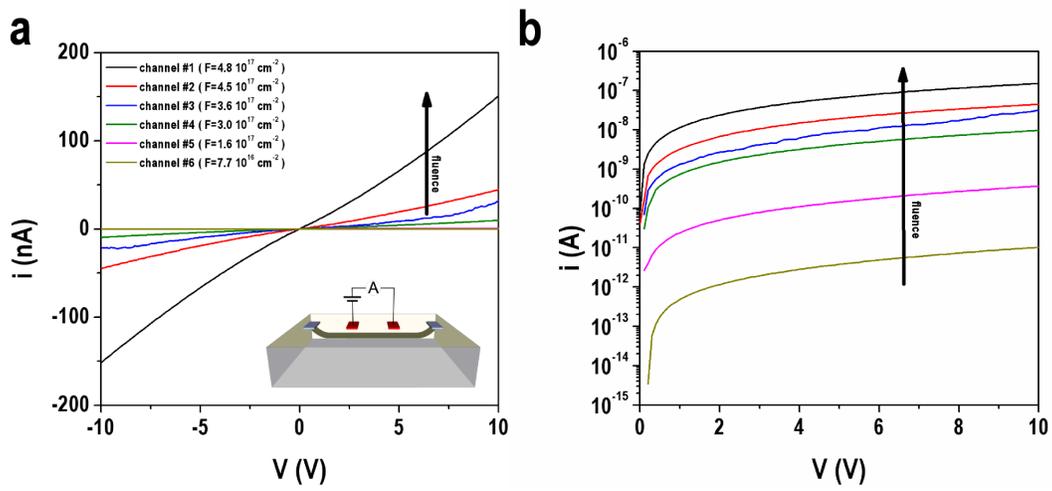

Figure 7

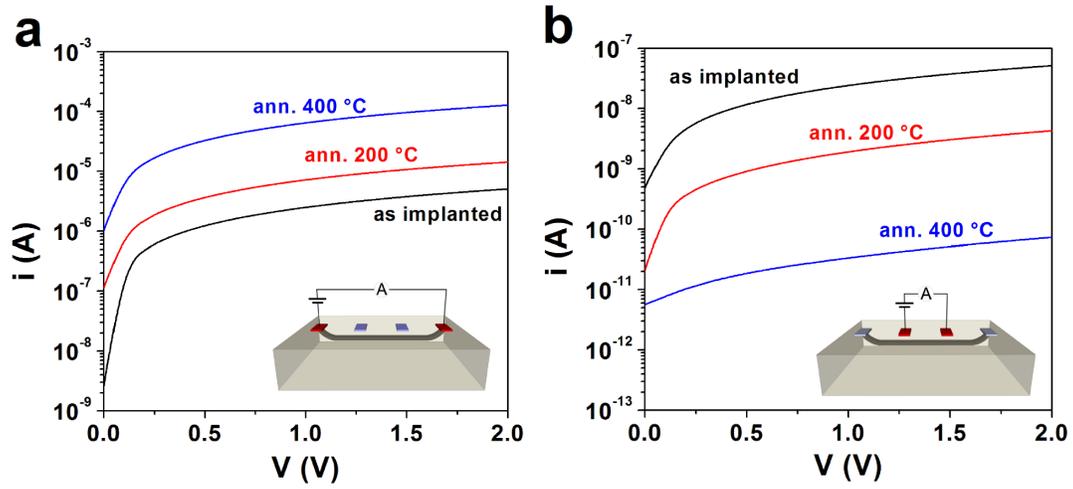

Figure 8

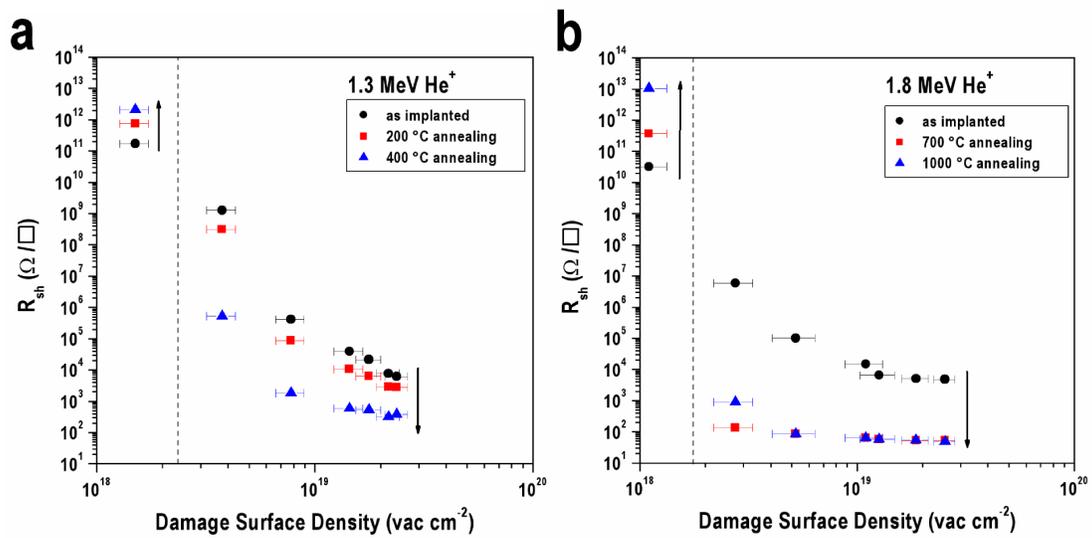

Figure 9

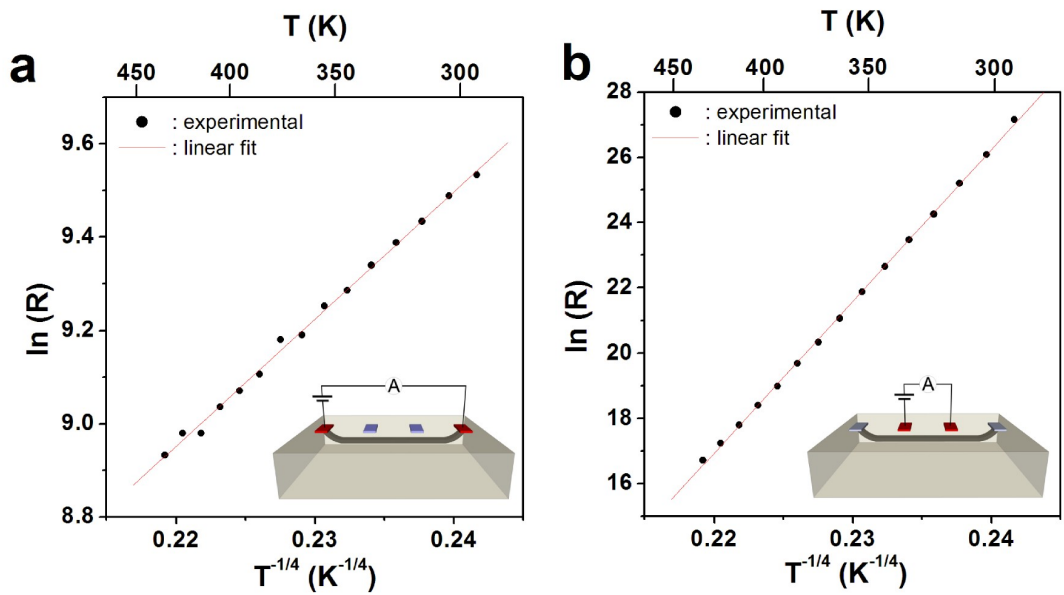

Figure 10

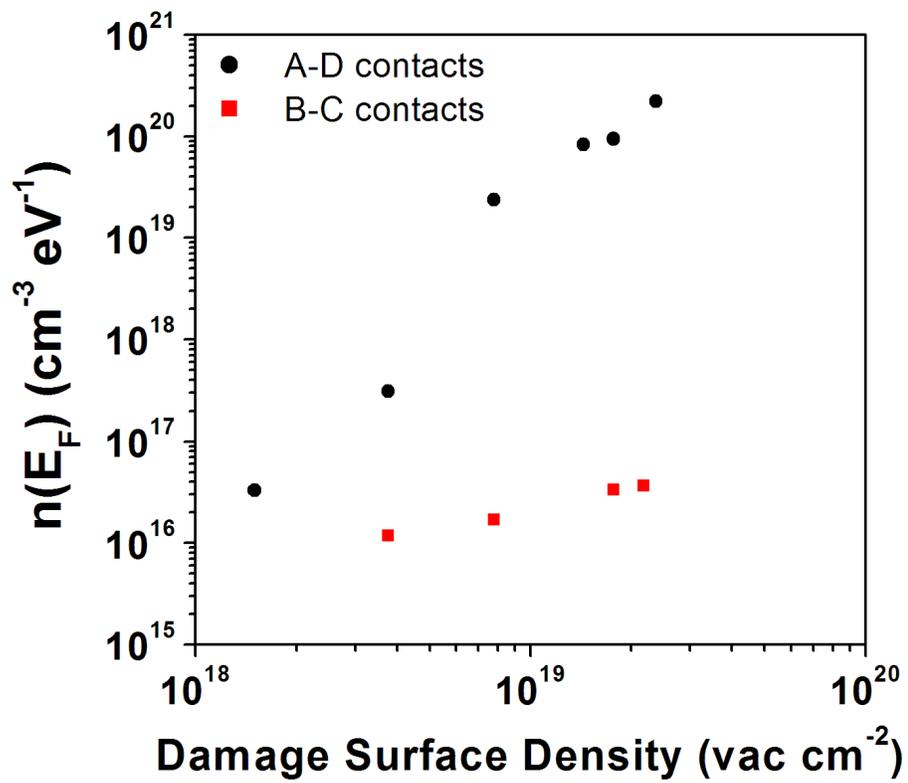

Figure 11

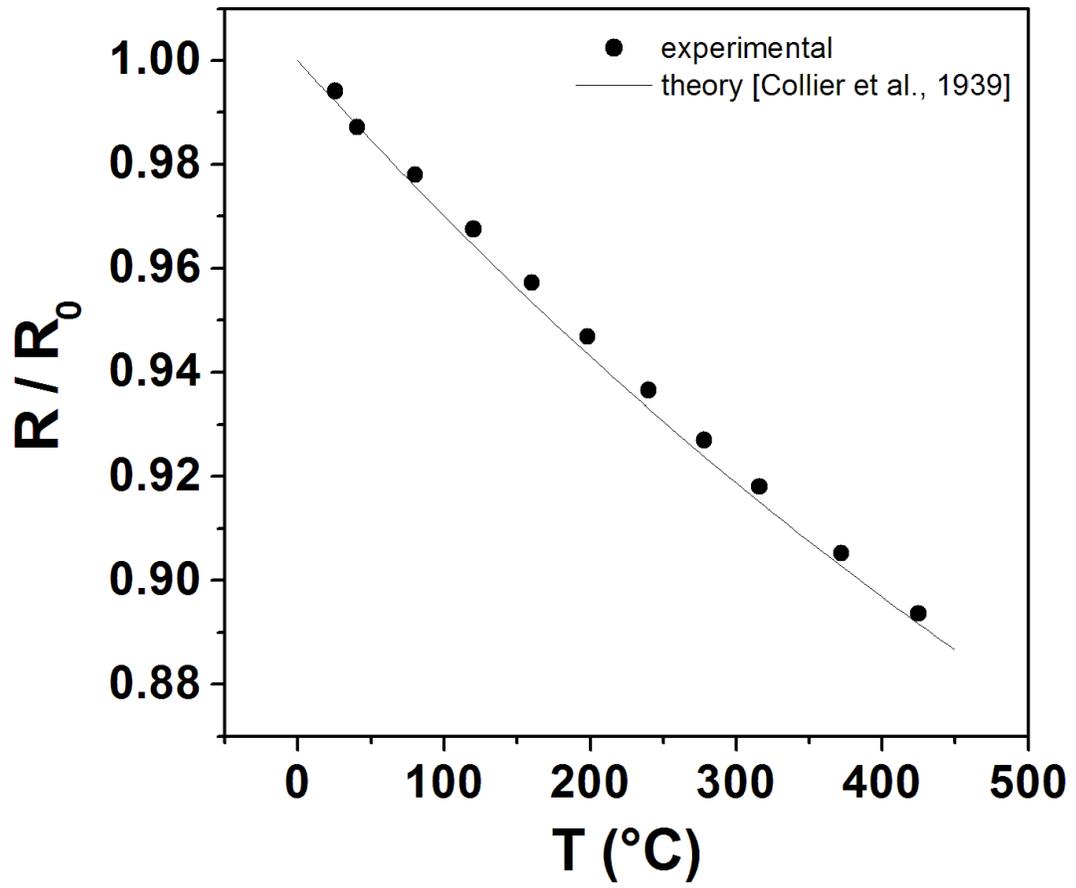

**Figure captions**

Figure 1 (color online): SRIM-2008.04 Monte Carlo simulations of the damage density profile induced in diamond by He ions with energies 0.5 MeV (blue plot), 1.3 MeV (red plot) and 1.8 MeV (green plot); the total number of vacancies/ion is 46, 47 and 52 for the three ion energies, respectively; the end-of-range damage peaks are clearly visible for all ion energies.

Figure 2 (color online): (a, b) Three-dimensional schematics (not in scale) of the deposition method of the variable-thickness metal mask: (a) the evaporation through a non-contact aperture defines metals masks with slowly degrading edges, while (b) MeV ion implantation across such edges determines a suitable modulation of the depth at which the highly damaged layer is formed. (c) Profilometry scan of a 5-µm-thick variable-thickness mask for 1.8 MeV He implantation.

Figure 3: TEM cross-sectional image of sample 1 after implantation across the sloping edge of a variable thickness mask; the formation of a buried amorphized layer and its progressive emergence towards the right-hand side of the image are clearly visible.

Figure 4 (color online): (a) three-dimensional schematics (not in scale) of the formation of emerging microchannels upon MeV ion implantation through a system of variable- thickness masks; (b) optical image in transmission of two typical microchannels after implantation in sample 2 at fluences $7.7 \cdot 10^{16}$ cm$^{-2}$ and $1.6 \cdot 10^{17}$ cm$^{-2}$ followed by mask removal; the buried channels are clearly visible since they are optically opaque; (c) three-dimensional schematics (not in scale) of the contacting of the sample with four Ag pads: contacts labeled as "A" and "D" are in direct electrical connection with the buried layer through its emerging endpoints, while "B" and "C" are defined between "A" and "D", above the buried channel.

Figure 5 (color online): current-voltage characteristics at room temperature of as-implanted channels in sample 2; the current values on the vertical axis are reported in linear scale in (a) to show the linear trend of the curves, while in (b) the logarithmic scale clearly highlights the increasing current values for channels implanted at higher fluences; the inset of (a) schematically reports the connection of the probing system to the A-D electrical contacts (see figure 4c).

Figure 6 (color online): current-voltage characteristics at room temperature of as-implanted channels in sample 2; the current values on the vertical axis are reported in linear scale in (a) to show the linear trend of the curves, while in (b) the logarithmic scale clearly highlights the increasing current values for channels implanted at higher fluences; the inset of (a) schematically reports the connection of the probing system to the B-C electrical contacts (see figure 4c).

Figure 7 (color online): current-voltage characteristics at room temperature of the channel implanted at fluence $F=4.5 \cdot 10^{17}$ cm$^{-2}$ in sample 2 implanted with 1.3 MeV He$^+$ ions, at various stages of the post-implantation thermal processing; data are reported for two probing contacts geometries (see figure 4c): A-D contacts in (a) and B-C contacts in (b), exhibiting opposite trends of the change in conductance as a function of annealing.

Figure 8 (color online): sheet resistance of microchannels measured between contacts A and D (see figure 4c) as a function of damage surface density, as measured after different stages of post-implantation annealing. The data are reported for samples 2 (a) and 3 (b), which were implanted with He$^+$ ions with energies of 1.3 MeV and 1.8 MeV, respectively. In both cases, the different effect of thermal annealing on the change in resistance is observed for different implantation fluences.

Figure 9 (color online): plots of $ln(R)$ versus $T^{-1/4}$ for the channel implanted at fluence $4.8 \cdot 10^{17}$ cm$^{-2}$ in sample 2, as measured both between contacts A-D (a) and B-D (b) (see also the schematics in the insets) after thermal annealing at 200 °C. Compatibly with the three-dimensional variable range hopping model, the experimental data (black dots) exhibit a linear trend; the relevant linear fits are reported in the red lines.

Figure 10 (color online): density of localized states at the Fermi level per unit of volume and energy as a function of the damage surface density resulting from the I-V characterization in temperature of sample 2 after 200 °C annealing. Data are reported for both the buried layer (black dots) and for the "cap layer" (red squares), which were probed in the A-D and B-C contact configurations respectively (see figure 4c).

Figure 11: normalized resistivity of the channel implanted at fluence $2.5 \cdot 10^{17}$ cm$^{-2}$ in sample 3 after 1000 °C thermal annealing as a function of measurement temperature (dots), and theoretical curve based on the phenomenological model reported for common graphitic carbon in [Collier1939].